\documentclass[preprintnumbers,amsmath,amssymb,nofootinbib]{revtex4}

\usepackage{graphicx}
\usepackage{epsfig}
\usepackage{color}

\newcommand{\beq}{\begin{equation}}
\newcommand{\eeq}{\end{equation}}
\newcommand{\bea}{\begin{eqnarray}}
\newcommand{\eea}{\end{eqnarray}}
\newcommand{\bwd}{\begin{widetext}}
\newcommand{\ewd}{\end{widetext}}

\begin{document}

\title{A symplectic particle-in-cell model for space-charge beam dynamics simulation
}


\author{Ji Qiang}
\email{jqiang@lbl.gov}
\affiliation{Lawrence Berkeley National Laboratory, Berkeley, CA 94720, USA}

\begin{abstract}
Space-charge effects play an important role in high intensity particle accelerators
and were studied using a variety of macroparticle tracking
models.
In this paper, we propose a symplectic particle-in-cell (PIC) model and 
compare this model with a recently published symplectic gridless particle model
and a conventional nonsymplectic PIC model for long-term
space-charge simulation of a coasting beam.
Using the same step size and the same number of modes for the space-charge
Poisson solver, all three models show qualitatively similar
emittance growth evolutions and final phase space shapes in the
benchmark examples. Quantitatively,
 the symplectic PIC and the symplectic
gridless particle models agree with each other very well, while
the nonsymplectic PIC model yields different emittance growth
value.
Using finer step size, the emittance growth from the 
nonsymplectic PIC converges towards that from the symplectic PIC model. 
\end{abstract}

\maketitle

\section{Introduction}

The nonlinear space-charge effects from particle interactions
inside a charged particle beam play an important role in high-intensity 
accelerators.
A variety of models have been developed to simulate the space-charge
effects~\cite{friedman,takeda,machida,jones,track,impact,toutatis,galambos,qin0,franchetti03,impact-t,amundson,opal}.
Among those models, the particle-in-cell (PIC) model is probably the most
widely used model in the
accelerator community.
The particle-in-cell model is an efficient model to handle the space-charge effects self-consistently.
It uses a computational grid to obtain the charge density distribution from
a finite number of macroparticles and solves the Poisson equation on the grid at each 
time step. The computational cost is linearly 
proportional to the number of macroparticles, which makes the simulation fast for
many applications. However, those grid based, momentum conserved, PIC codes do not 
satisfy the symplectic condition of classic multi-particle dynamics. 
Violating the symplectic condition
in multi-particle tracking might not be an issue in a single pass system such as
a linear accelerator. 
But in a circular accelerator, 
violating the symplectic condition
may result in undesired numerical errors in the long-term tracking simulation.
Recently, a symplectic gridless particle model was proposed for
self-consistent space-charge simulation~\cite{qiang17}. This model is 
easy to implement and has a good scalability on massive parallel computers.
However, the computational cost of this model is proportional to the 
number of macroparticles times the number of modes used to solve the Poisson
equation. When the number of modes is large, this model becomes quite
computationally expensive and has to resort to parallel computers.

In this paper, we propose a symplectic PIC model for self-consistent
space-charge long-term simulation. This model has the 
advantages of the computational efficiency of the PIC method and the
symplectic property needed for long-term dynamics tracking. 
Multi-symplectic particle-in-cell model was
proposed to study Vlasov-Maxwell system in plasmas using
a variational method~\cite{xiao,evstatiev,shadwick,qin}.
A symplectic gridless spectral electrostatic model in a periodic plasma was
also proposed following the variational method~\cite{webb}. 
To the best of our knowledge, at present,
there is no symplectic PIC model that was proposed and shown
applications for the quasi-static
Vlasov-Poisson system in space-charge beam dynamics study.
In this paper, we derived a symplectic PIC model directly from
the multi-particle Hamiltonian. We then carried out long-term
space-charge simulations and compared this 
model with the symplectic gridless particle model and the nonsymplectic
PIC model in two beam dynamics applications.

The organization of this paper is as follows: after the Introduction, we present the
symplectic multi-particle tracking model including the space-charge effect in Section II; 
We present
the nonsymplectic PIC model in Section III; We compare the symplectic PIC model,
the symplectic gridless particle model, and the nonsymplectic PIC model with two
space-charge simulation examples in Section IV;
and draw conclusions in Section V.

\section{Symplectic Multi-Particle Tracking Models}

In the accelerator beam dynamics simulation, for a multi-particle system with $N_p$ charged particles subject to both a space-charge self field
and an external field, an approximate Hamiltonian of the system can 
be written as~\cite{rob1,forest,alex}:
\begin{eqnarray}
	H & = & \sum_{i=1}^{N_p} {\bf p}_i^2/2 + \frac{1}{2} \sum_{i=1}^{N_p} 
	\sum_{j=1}^{N_p} q \varphi({\bf r}_i,{\bf r}_j)
	+ \sum_{i=1}^{N_p} q \psi({\bf r}_i)
	\label{htot}
\end{eqnarray}
where $H({\bf r}_1,{\bf r}_2,\cdots,{\bf r}_{N_p},{\bf p}_1,{\bf p}_2,\cdots,{\bf p}_{N_p};s)$ denotes the Hamiltonian of the system using distance $s$ as an independent variable, 
$\varphi$ is related to the space-charge interaction potential 
between the charged
particles $i$ and $j$ (subject to appropriate boundary conditions), $\psi$ denotes the
potential associated with the external field,
${\bf r}_i=(x_i,y_i,\theta_i=\omega \Delta t)$ denotes the normalized canonical
spatial coordinates of particle $i$, ${\bf p}_i=(p_{xi},p_{yi},p_{ti}=-\Delta E/mc^2)$ the normalized canonical momentum
coordinates of particle $i$, and $\omega$ the reference angular frequency,
$\Delta t$ the time of flight to location $s$, $\Delta E$ the energy
deviation with respect to the reference particle, $m$ the rest mass of the
particle, and $c$ the speed of light in vacuum.
The equations governing the motion of individual particle
$i$ follow the Hamilton's equations as:
\begin{eqnarray}
	\frac{d {\bf r}_i}{d s} & = & \frac{\partial H}{\partial {\bf p}_i} \\
	\frac{d {\bf p}_i}{d s} & = & -\frac{\partial H}{\partial {\bf r}_i} 
\end{eqnarray}
Let $\zeta$ denote a 6N-vector of coordinates,
the above Hamilton's equation can be rewritten as:
\begin{eqnarray}
	\frac{d \zeta}{d s} & = & -[H, \zeta] 
\end{eqnarray}
where [\ ,\ ] is the Poisson bracket. A formal solution for above equation
after a single step $\tau$ can be written as:
\begin{eqnarray}
	\zeta (\tau) & = & \exp(-\tau(:H:)) \zeta(0)
\end{eqnarray}
Here, we have defined a differential operator $:H:$ as $: H : g = [H, \ g]$, 
for arbitrary function $g$. 
For a Hamiltonian that can be written as a sum of two terms $H =  H_1 + H_2$, an approximate
solution to above formal solution can be written as~\cite{forest1}
\begin{eqnarray}
	\zeta (\tau) & = & \exp(-\tau(:H_1:+:H_2:)) \zeta(0) \nonumber \\
  & = & \exp(-\frac{1}{2}\tau :H_1:)\exp(-\tau:H_2:) \exp(-\frac{1}{2}\tau:H_1:) \zeta(0) + O(\tau^3)
\end{eqnarray}
Let $\exp(-\frac{1}{2}\tau :H_1:)$ define a transfer map ${\mathcal M}_1$ and
$\exp(-\tau:H_2:)$ a transfer map ${\mathcal M}_2$, 
for a single step, the above splitting results in a second order numerical integrator
for the original Hamilton's equation as:
\begin{eqnarray}
	\zeta (\tau) & = & {\mathcal M}(\tau) \zeta(0) \nonumber \\
    & = & {\mathcal M}_1(\tau/2) {\mathcal M}_2(\tau) {\mathcal M}_1(\tau/2) \zeta(0)
	+ O(\tau^3)
	\label{map}
\end{eqnarray}

The above numerical integrator can be extended to $4^{th}$ order 
accuracy and arbitrary even-order accuracy following
Yoshida's approach~\cite{yoshida}.
This numerical integrator Eq.~\ref{map} will be symplectic if both the transfer map
${\mathcal M}_1$ and the transfer map ${\mathcal M}_2$ are symplectic.
A transfer map ${\mathcal M}_i$ is symplectic if and only if
the Jacobian matrix $M_i$ of the transfer
map ${\mathcal M}_i$ satisfies the following condition:
\begin{eqnarray}
M_i^T J M_i = J 
\label{symp}
\end{eqnarray}
where $J$ denotes the $6N \times 6N$ matrix given by:
\begin{equation}
	J  =   \left( \begin{array}{cc}
			0 & I \\
			-I & 0
	\end{array} \right)
\end{equation}
and $I$ is the $3N\times 3N$ identity matrix.

For the Hamiltonian in Eq.~\ref{htot}, we can choose $H_1$ as:
\begin{eqnarray}
	H_1 & = & \sum_{i=1}^{N_p} {\bf p}_i^2/2 + \sum_{i=1}^{N_p} q \psi({\bf r}_i)
\end{eqnarray}
This corresponds to the Hamiltonian of a group of charged particles 
inside an external field without mutual interaction among themselves. 
A single-charged particle magnetic optics method can be used to find 
the symplectic transfer map ${\mathcal M}_1$
for this Hamiltonian with the external fields from
most accelerator beam line elements~\cite{rob1,alex,mad}.


We can choose $H_2$ as:
\begin{eqnarray}
	H_2 & = & \frac{1}{2} \sum_{i=1}^{N_p} \sum_{j=1}^{N_p} q \varphi({\bf r}_i,{\bf r}_j)
\end{eqnarray}
which includes the space-charge effect and is only a function of position. 
For the space-charge Hamiltonian $H_2({\bf r})$, the single
step transfer map ${\mathcal M}_2$ can be written as:
\begin{eqnarray}
	{\bf r}_i(\tau) & = & {\bf r}_i(0) \\
	{\bf p}_i(\tau) & = & {\bf p}_i(0) - \frac{\partial H_2({\bf r})}{\partial {\bf r}_i} \tau
	\label{map2}
\end{eqnarray}
The Jacobi matrix of the above 
 transfer map
${\mathcal M}_2$ is 
\begin{equation}
	M_2  =   \left( \begin{array}{cc}
			I & 0 \\
			L & I
	\end{array} \right)
\end{equation}
where $L$ is a $3N \times 3N$ matrix.
For $M_2$ to satisfy the symplectic condition Eq.~\ref{symp}, 
the matrix $L$ needs
to be a symmetric matrix, i.e.
\begin{equation}
	L = L^T
\end{equation}
Given the fact that $L_{ij} = \partial {\bf p}_i(\tau)/\partial {\bf r}_j =  - \frac{\partial^2 H_2({\bf r})}{\partial {\bf r}_i \partial {\bf r}_j} \tau $, the matrix $L$ will be symmetric as long as it 
is {\bf \it analytically calculated}
from the function $H_2$. This is also called jolt-factorization in nonlinear
single particle beam dynamics study~\cite{forest3}.
If both the transfer map
${\mathcal M}_1$ and the transfer map ${\mathcal M}_2$ 
are symplectic, the numerical integrator Eq.~\ref{map} for multi-particle tracking will be symplectic. 


For a coasting beam, the Hamiltonian $H_2$ can be written as~\cite{rob1}:
\begin{eqnarray}
	H_2 & = & \frac{K}{2} \sum_{i=1}^{N_p} \sum_{j=1}^{N_p} \varphi({\bf r}_i,{\bf r}_j)
	\label{htot2}
\end{eqnarray}
where $K = q I/(2\pi \epsilon_0 p_0 v_0^2 \gamma_0^2)$ is the generalized perveance,
$I$ is the beam current, $\epsilon_0$ is the permittivity of vacuum,
$p_0$ is the momentum of the reference
particle, $v_0$ is the speed of the reference particle, $\gamma_0$ is
the relativistic factor of the reference particle, and $\varphi$ is the 
space charge Coulomb interaction potential.
In this Hamiltonian, the effects of the direct Coulomb 
electric potential and the
longitudinal vector potential are combined together.
The electric Coulomb potential in the Hamiltonian $H_2$ can be obtained 
from the solution of the Poisson equation.
In the following, we assume that the coasting beam is inside a rectangular perfectly conducting pipe.
In this case, the two-dimensional Poisson's equation can be written as:
\begin{equation}
\frac{\partial^2 \phi}{\partial x^2} +
\frac{\partial^2 \phi}{\partial y^2} = - 4 \pi \rho
\label{poi2d}
\end{equation}
where
$\phi$ is the electric potential, and $\rho$ is the particle
density distribution of the beam.

The boundary conditions for the electric potential inside the rectangular 
perfectly conducting pipe are:
\begin{eqnarray}
	\label{bc1}
\phi(x=0,y) & = & 0  \\
\phi(x=a,y) & = & 0  \\
\phi(x,y=0) & = & 0  \\
\phi(x,y=b) & = & 0  
	\label{boundary}
\end{eqnarray}
where $a$ is the horizontal width of the pipe and $b$ is the vertical width
of the pipe. 

Given the boundary conditions in Eqs.~\ref{bc1}-\ref{boundary}, the electric potential $\phi$ and the
source term $\rho$ can be approximated using two sine functions as~\cite{gottlieb,fornberg,boyd,qiang1,qiang2}:
\begin{eqnarray}
	\rho(x,y)  = \sum_{l=1}^{N_l}\sum_{m=1}^{N_m} \rho^{lm} \sin(\alpha_l x) \sin(\beta_m y) \\
	\phi(x,y)  =  \sum_{l=1}^{N_l}\sum_{m=1}^{N_m} \phi^{lm} \sin(\alpha_l x) \sin(\beta_m y) 
\end{eqnarray}
where
\begin{eqnarray}
\label{rholm}
\rho^{lm}  = \frac{4}{ab}\int_0^a\int_0^b \rho(x,y) \sin(\alpha_l x) \sin(\beta_m y) \ dx dy \\
\phi^{lm}  = \frac{4}{ab}\int_0^a\int_0^b \phi(x,y) \sin(\alpha_l x) \sin(\beta_m y) \ dx dy
\end{eqnarray}
where $\alpha_l=l\pi/a$ and $\beta_m = m \pi/b$.
The above approximation
follows the numerical spectral Galerkin method since each basis function
satisfies the boundary conditions on the wall~\cite{gottlieb,boyd,fornberg}. 
For a smooth function,
this spectral approximation has an accuracy whose numerical error
scales as $O(\exp(-cN))$ with 
$c>0$, where $N$ is the number of the basis function (i.e. mode number in each
dimension) used in the approximation.
By substituting above expansions into the
Poisson Eq.~\ref{poi2d} and making use of the orthonormal condition of the sine functions,
we obtain
\begin{eqnarray}
	\phi^{lm} & = & \frac{4 \pi \rho^{lm}}{\gamma_{lm}^2}
	\label{odelm}
\end{eqnarray}
where $\gamma_{lm}^2 = \alpha_l^2 + \beta_m^2$. 

In the simulation, the particle distribution function $\rho(x,y)$ can be 
represented as:
\begin{eqnarray}
	\rho(x,y) & = & \frac{1}{N_p}\sum_{j=1}^{N_p} S(x-x_j) S(y-y_j)
\end{eqnarray}
where 
$S(x)$ is the shape function (also called deposition function in the PIC model).
Using the above equation and Eq.~\ref{rholm} and Eq.~\ref{odelm}, we obtain:
\begin{eqnarray}
	\phi^{lm} &  = & \frac{4 \pi} {\gamma_{lm}^2}\frac{4}{ab} \frac{1}{N_p} \sum_{j=1}^{N_p} 
	\int_0^a\int_0^b S(x-x_j)S(y-y_j) \sin(\alpha_l x) \sin(\beta_m y) dxdy
\end{eqnarray}
and the electric potential as:
\begin{eqnarray}
	\phi(x,y)  = {4 \pi} \frac{4}{ab} \frac{1}{N_p} \sum_{j=1}^{N_p} 
	\sum_{l=1}^{N_l} \sum_{m=1}^{N_m} 
	\frac{1}{\gamma_{lm}^2} \sin(\alpha_l x) \sin(\beta_m y) 
	\int_0^a\int_0^b S(x-x_j)S(y-y_j) \sin(\alpha_l x) \sin(\beta_m y) dxdy
\end{eqnarray}
The electric potential at a particle $i$ location can be obtained from the
potential on grid as:
\begin{eqnarray}
	\phi(x_i,y_i) & = & \int_0^a \int_0^b \phi(x,y) S(x-x_i)S(y-y_i) dxdy
\end{eqnarray}
where the interpolation function from the grid to the particle location is
assumed to be the same as the deposition function.

From the above electric potential, the interaction potential 
$\varphi$ between particles $i$ and $j$ can be
written as:
\begin{eqnarray}
	\varphi(x_i,y_i,x_j,y_j) & = & {4 \pi} \frac{4}{ab} \frac{1}{N_p}  
	\sum_{l=1}^{N_l} \sum_{m=1}^{N_m} 
	\frac{1}{\gamma_{lm}^2} 
	\int_0^a\int_0^b S(x-x_j)S(y-y_j) \sin(\alpha_l x) \sin(\beta_m y) dxdy \nonumber \\
	& &	\int_0^a \int_0^b S(x-x_i)S(y-y_i) \sin(\alpha_l x) \sin(\beta_m y) dxdy
\end{eqnarray}
Now, the space-charge Hamiltonian $H_2$ can be written as:
\begin{eqnarray}
	H_2 & = & 4\pi \frac{K}{2}\frac{4}{ab} \frac{1}{N_p} \sum_{i=1}^{N_p} \sum_{j=1}^{N_p} \sum_{l=1}^{N_l} \sum_{m=1}^{N_m} 
	\frac{1}{\gamma_{lm}^2}  \nonumber \\
	& &	\int_0^a\int_0^b S(x-x_j)S(y-y_j) \sin(\alpha_l x) \sin(\beta_m y) dxdy
		\int_0^a \int_0^b S(x-x_i)S(y-y_i) \sin(\alpha_l x) \sin(\beta_m y) dxdy
\end{eqnarray}

\subsection{Symplectic Gridless Particle Model}

In the symplectic gridless particle space-charge model, 
the shape function is assumed to be a Dirac delta function and the
particle distribution function $\rho(x,y)$ can be represented as:
\begin{eqnarray}
	\rho(x,y) = \frac{1}{N_p}\sum_{j=1}^{N_p} \delta(x-x_j)\delta(y-y_j)
\end{eqnarray}


Now, the space-charge Hamiltonian $H_2$ can be written as:
\begin{eqnarray}
	H_2  = 4\pi \frac{K}{2}\frac{4}{ab} \frac{1}{N_p} \sum_{i=1}^{N_p} \sum_{j=1}^{N_p} \sum_{l=1}^{N_l} \sum_{m=1}^{N_m} 
	\frac{1}{\gamma_{lm}^2} \sin(\alpha_l x_j) 
	\sin(\beta_m y_j) \sin(\alpha_l x_i) \sin(\beta_m y_i)
\end{eqnarray}
The one-step symplectic transfer map ${\mathcal M}_2$ of the 
particle $i$ with this Hamiltonian is given as:
\begin{eqnarray}
	p_{xi}(\tau) & = & p_{xi}(0) -
	\tau 4\pi {K}\frac{4}{ab} \frac{1}{N_p} \sum_{j=1}^{N_p} 
	\sum_{l=1}^{N_l} \sum_{m=1}^{N_m} 
	\frac{\alpha_l}{\gamma_{lm}^2} 
 \sin(\alpha_l x_j) \sin(\beta_m y_j) \cos(\alpha_l x_i) \sin(\beta_m y_i)  \nonumber \\
	p_{yi}(\tau) & = & p_{yi}(0) -
	\tau 4\pi {K}\frac{4}{ab} \frac{1}{N_p} \sum_{j=1}^{N_p} 
	\sum_{l=1}^{N_l} \sum_{m=1}^{N_m} 
	\frac{\beta_m}{\gamma_{lm}^2} 
 \sin(\alpha_l x_j) \sin(\beta_m y_j) \sin(\alpha_l x_i) \cos(\beta_m y_i)
\end{eqnarray}
Here, both $p_{xi}$ and $p_{yi}$ are normalized by the reference particle momentum $p_0$.

\subsection{Symplectic Particle-In-Cell Model}

If the deposition/interpolation shape function is not a delta function,
the one-step symplectic transfer map ${\mathcal M}_2$ of the 
particle $i$ with this space-charge Hamiltonian $H_2$ is given as:
\begin{eqnarray}
	p_{xi}(\tau) & = & p_{xi}(0) -
	\tau 4\pi {K}\frac{4}{ab} \frac{1}{N_p} \sum_{j=1}^{N_p} 
	\sum_{l=1}^{N_l} \sum_{m=1}^{N_m} 
	\frac{1}{\gamma_{lm}^2} 
	\int_0^a\int_0^b S(x-x_j)S(y-y_j) \sin(\alpha_l x) \sin(\beta_m y) dxdy
     \nonumber \\ 
	& &	\int_0^a \int_0^b 
	\frac{\partial S(x-x_i)}{\partial x_i}S(y-y_i) \sin(\alpha_l x) \sin(\beta_m y) dxdy
	\nonumber \\
	p_{yi}(\tau) & = & p_{yi}(0) -
	\tau 4\pi {K}\frac{4}{ab} \frac{1}{N_p} \sum_{j=1}^{N_p} 
	\sum_{l=1}^{N_l} \sum_{m=1}^{N_m} 
	\frac{1}{\gamma_{lm}^2} 
	\int_0^a\int_0^b S(x-x_j)S(y-y_j) \sin(\alpha_l x) \sin(\beta_m y) dxdy
     \nonumber \\ 
	& &	\int_0^a \int_0^b S(x-x_i)\frac{\partial S(y-y_i)}{\partial y_i} \sin(\alpha_l x) \sin(\beta_m y) dxdy
\end{eqnarray}
where both $p_{xi}$ and $p_{yi}$ are normalized by the reference particle momentum $p_0$,
and $S'(x)$ is the first derivative of the shape function.
Assuming that the shape function is a compact local function with respect to
the computational grid, the integral in the above map can be approximated as:
\begin{eqnarray}
	p_{xi}(\tau) & = & p_{xi}(0) -
	\tau 4\pi {K}\frac{4}{ab} \frac{1}{N_p} \sum_{j=1}^{N_p} 
	\sum_{l=1}^{N_l} \sum_{m=1}^{N_m} 
	\frac{1}{\gamma_{lm}^2} 
	\sum_{I'} \sum_{J'} S(x_{I'}-x_j)S(y_{J'}-y_j) \sin(\alpha_l x_{I'}) \sin(\beta_m y_{J'}) 
     \nonumber \\ 
	& &	\sum_I \sum_J \frac{\partial S(x_I-x_i)}{\partial x_i}S(y_J-y_i) \sin(\alpha_l x_I) \sin(\beta_m y_J) 
	\nonumber \\
	p_{yi}(\tau) & = & p_{yi}(0) -
	\tau 4\pi {K}\frac{4}{ab} \frac{1}{N_p} \sum_{j=1}^{N_p} 
	\sum_{l=1}^{N_l} \sum_{m=1}^{N_m} 
	\frac{1}{\gamma_{lm}^2} 
	\sum_{I'} \sum_{J'} S(x_{I'}-x_j)S(y_{J'}-y_j) \sin(\alpha_l x_{I'}) \sin(\beta_m y_{J'})
     \nonumber \\ 
	& &	\sum_I \sum_J S(x_I-x_i)\frac{\partial S(y_I-y_i)}{\partial y_i} \sin(\alpha_l x_I) \sin(\beta_m y_J) 
\end{eqnarray}
where the integers $I$, $J$, $I'$, and $J'$ denote the two dimensional computational
grid index, and the summations with respect to those indices are limited to the range
of a few local grid points depending on the specific deposition function.
If one defines the charge density $\rho(x_I',y_J')$ on the grid as:
\begin{eqnarray}
	\rho(x_{I'},y_{J'})  = \frac{1}{N_p} \sum_{j=1}^{N_p}  S(x_{I'}-x_j)S(y_{J'}-y_j),
	\label{2dep}
\end{eqnarray}
the space-charge map can be rewritten as:
\begin{eqnarray}
		p_{xi}(\tau) & = & p_{xi}(0) -
	\tau 4\pi {K}\sum_I \sum_J \frac{\partial S(x_I-x_i)}{\partial x_i} S(y_J-y_i) [\frac{4}{ab} 
	\sum_{l=1}^{N_l} \sum_{m=1}^{N_m} 
	\frac{1}{\gamma_{lm}^2} 
	\nonumber \\
	& & \sum_{I'} \sum_{J'} \rho(x_{I'},y_{J'}) \sin(\alpha_l x_{I'}) \sin(\beta_m y_{J'}) 
	 \sin(\alpha_l x_I) \sin(\beta_m y_J) ]
	\nonumber \\
	p_{yi}(\tau) & = & p_{yi}(0) -
	\tau 4\pi {K}\sum_I \sum_J S(x_I-x_i)\frac{\partial S(y_I-y_i)}{\partial y_i} [\frac{4}{ab}   
	\sum_{l=1}^{N_l} \sum_{m=1}^{N_m} 
	\frac{1}{\gamma_{lm}^2} 
	\nonumber \\
	& & \sum_{I'} \sum_{J'}	\rho(x_{I'},y_{J'}) \sin(\alpha_l x_{I'}) \sin(\beta_m y_{J'})
	\sin(\alpha_l x_I) \sin(\beta_m y_J) ]
\end{eqnarray}
It turns out that the expression inside the bracket represents the solution of potential
on grid using the charge density on grid, which can be written as:
\begin{eqnarray}
 \phi(x_I,y_J) & = & \frac{4}{ab} 
	\sum_{l=1}^{N_l} \sum_{m=1}^{N_m} 
	\frac{1}{\gamma_{lm}^2} 
\sum_{I'} \sum_{J'} \rho(x_{I'},y_{J'}) \sin(\alpha_l x_{I'}) \sin(\beta_m y_{J'}) 
	 \sin(\alpha_l x_I) \sin(\beta_m y_J) 
	 \label{poi2dsol}
\end{eqnarray}
Then the above space-charge map can be rewritten in more concise format as:
\begin{eqnarray}
		p_{xi}(\tau) & = & p_{xi}(0) -
	\tau 4\pi {K}\sum_I \sum_J \frac{\partial S(x_I-x_i)}{\partial x_i} S(y_J-y_i) 
\phi(x_I,y_J)	\nonumber \\
	p_{yi}(\tau) & = & p_{yi}(0) -
	\tau 4\pi {K}\sum_I \sum_J S(x_I-x_i)\frac{\partial S(y_J-y_i)}{\partial y_i} 
	\phi(x_I,y_J)
\end{eqnarray}
In the PIC literature, some compact function such as linear function and quadratic function
is used in the simulation. For example, a quadratic shape function can be written as~\cite{hockney,birdsall}:
\begin{eqnarray}
	S(x_I-x_i) & = & \left \{ \begin{array}{ll} 
		\frac{3}{4} - (\frac{x_i - x_I}{h})^2, & |x_i - x_I| \leq h/2  \\
 \frac{1}{2}(\frac{3}{2}-\frac{|x_i-x_I|}{h})^2, & h/2 < |x_i - x_I| \leq 3/2 h \\
		 0 & {\rm otherwise}
                    \end{array}
                  \right.
\end{eqnarray}
\begin{eqnarray}
	\frac{\partial S(x_I-x_i)}{\partial x_i} & = & \left \{ \begin{array}{ll} 
		- 2(\frac{x_i - x_I}{h})/h, & |x_i - x_I| \leq h/2  \\
		(-\frac{3}{2}+\frac{(x_i-x_I)}{h})/h, & h/2 < |x_i - x_I| \leq 3/2 h, \ \  x_i > x_I \\
		(\frac{3}{2}+\frac{(x_i-x_I)}{h})/h, & h/2 < |x_i - x_I| \leq 3/2 h, \ \   x_i \leq x_I \\
		 0 & {\rm otherwise}
                    \end{array}
                  \right.
\end{eqnarray}
The same shape function and its derivative can be applied to the $y$ dimension. 

Using the symplectic transfer map ${\mathcal M}_1$ for the external field Hamiltonian 
$H_1$ from a magnetic optics code and the transfer map ${\mathcal M}_2$ for
space-charge Hamiltonian $H_2$,
one obtains a 
symplecti PIC model including the self-consistent space-charge effects.

\section{Nonsymplectic Particle-In-Cell Model}

In the nonsymplectic PIC model, for a coasting beam,
the equations governing the motion of individual particle
$i$ follow the Lorentz equations as:
\begin{eqnarray}
	\frac{d {\bf r}_i}{d s} & = & {\bf p}_i \\
	\frac{d {\bf p}_i}{d s} & = & q({\bf E}_i/v_0 - a_z \times {\bf B}_i) 
\end{eqnarray}
where ${\bf r} = (x,y)$, ${\bf p} = (v_x/v_0,v_y/v_0)$, 
and $a_z$ is a unit vector in the longitudinal $z$ direction.

The above equations can be solved using a $2^{nd}$ order leap-frog algorithm, 
which
is widely used in many dynamics simulations. Here, for each step, the
positions of a particle $i$ are advanced a half step by:
\begin{eqnarray}
	{\bf r}(\tau/2)_i & = & {\bf r}(0)_i + \frac{1}{2} \tau {\bf p}_i(0)
\end{eqnarray}
Then the momenta of the particle are advanced a full step
including the forces from the given external fields and the 
self-consistent space-charge fields.
In the PIC model,
the space-charge fields are obtained by solving the Poisson equation
on a computational grid and then interpolating back to
the particle location. In this study, we employ the above spectral
method to attain the space-charge fields as:
\begin{eqnarray}
 E_x(x_I,y_J) & = & -\frac{4}{ab} 
	\sum_{l=1}^{N_l} \sum_{m=1}^{N_m} 
	\frac{\alpha_l}{\gamma_{lm}^2} 
\sum_{I'} \sum_{J'} \rho(x_{I'},y_{J'}) \sin(\alpha_l x_{I'}) \sin(\beta_m y_{J'}) 
	 \cos(\alpha_l x_I) \sin(\beta_m y_J)  \\
 E_y(x_I,y_J) & = & -\frac{4}{ab} 
	\sum_{l=1}^{N_l} \sum_{m=1}^{N_m} 
	\frac{\beta_m}{\gamma_{lm}^2} 
\sum_{I'} \sum_{J'} \rho(x_{I'},y_{J'}) \sin(\alpha_l x_{I'}) \sin(\beta_m y_{J'}) 
	 \sin(\alpha_l x_I) \cos(\beta_m y_J)  
	 \label{exey}
\end{eqnarray}
where the charge density $\rho(x_{I'},y_{J'})$ on the grid is obtained 
following the charge deposition Eq.~\ref{2dep}. 
The particle momenta can then be advanced one step using:
\begin{eqnarray}
	p_{xi}(\tau) & = & p_{xi}(0)+\tau(\frac{q E^{ext}_x}{v_0}-q B^{ext}_y)
+ \tau 4\pi {K}\sum_I \sum_J S(x_I-x_i) S(y_J-y_i) 
E_x(x_I,y_J)	\nonumber \\
	p_{yi}(\tau) & = & p_{yi}(0)+\tau(\frac{q E^{ext}_y}{v_0}+q B^{ext}_x) + 
	\tau 4\pi {K}\sum_I \sum_J S(x_I-x_i) S(y_J-y_i) 
	E_y(x_I,y_J)
\end{eqnarray}
Here, both the direct electric Coulomb field and the magnetic field from
the longitudinal vector potential are included in the space-charge
force term of the above equation.
Then, the positions of the particle are advanced another half step using the new momenta as:
\begin{eqnarray}
	{\bf r}(\tau)_i & = & {\bf r}(\tau/2)_i + \frac{1}{2} \tau {\bf p}_i(\tau)
\end{eqnarray}

This procedure is repeated many steps during the simulation.

\section{Benchmark Examples}

In this section, we tested and compared the symplectic PIC model with
the symplectic gridless particle model and the nonsymplectic PIC model
using two application examples.
In both examples, the beam experienced
strong space-charge driven resonance. 
It was tracked including self-consistent
space-charge effects for several hundred thousands of
lattice periods using the above three models. 

\begin{figure}[!htb]
   \centering
   \includegraphics*[angle=0,width=230pt]{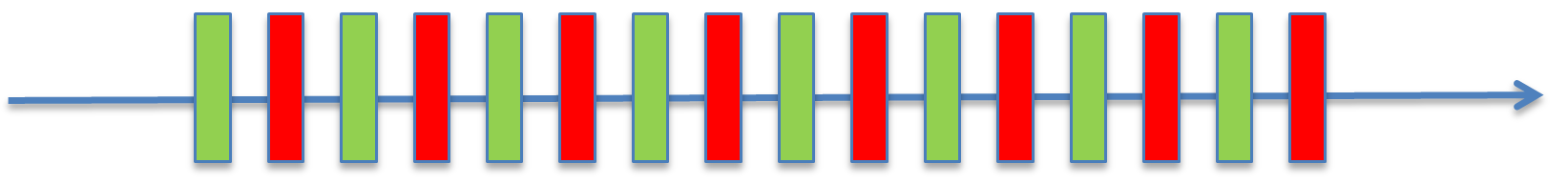}
   \caption{Schematic plot of a FODO lattice.}
   \label{fodo}
\end{figure}

In the first example, we simulated a $1$ GeV coasting proton beam 
transport through a 
linear periodic quadrupole focusing and defocusing (FODO) channel
inside a rectangular perfectly conducting pipe.
A schematic plot of the lattice is shown in Fig.~\ref{fodo}.
It consists of a $0.1$ m focusing quadrupole magnet and a $0.1$ m defocusing 
quadrupole magnet with a single period. 
The total length of the period is $1$ meter.
The zero current phase advance through one lattice period is $85$ degrees.
The current of the beam is $450$ A and the depressed phase advance
is $42$ degrees. Such a high current drives the beam across
the $4^{th}$ order collective instability~\cite{ingo83}.
The initial transverse normalized emittance of the proton beam is $1$ $\mu m$ 
with a 4D Gaussian distribution. 

\begin{figure}[!htb]
   \centering
   \includegraphics*[angle=0,width=230pt]{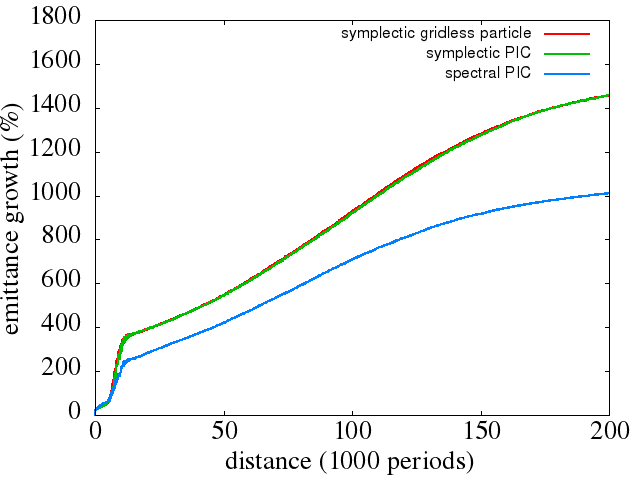}
   \caption{
	   Four dimensional emittance growth evolution from the
	symplectic gridless particle model (red), the symplectic PIC model (green),
	and the nonsymplectic spectral PIC (blue).}
   \label{emtfodo}
\end{figure}

Figure~\ref{emtfodo} shows the four dimensional emittance growth
$(\frac{\epsilon_x}{\epsilon_{x0}}\frac{\epsilon_y}{\epsilon_{y0}}-1)\%$
evolution through $200,000$ lattice periods from the symplectic gridless
particle model, from the symplectic PIC model, and from the nonsymplectic
spectral PIC model. These simulations used about $50,000$ macroparticles
and $15\times 15$ modes and $257\times 257$ grid points in the spectral Poisson solver.
The perfectly conducting pipe has a square shape with an aperture size
of $10$ mm.
It is seen that the symplectic PIC model and the symplectic gridless particle
model agrees with each other very well. The nonsymplectic spectral PIC model 
yields significantly
smaller emittance growth than those from the two symplectic methods,
which might result from the numerical damping effects in the nonsymplectic
integrator.
The fast emittance growth within the first $20,000$ periods is 
caused by the space-charge driven $4^{th}$ order collective instability.
The slow emittance growth after $20,000$ periods might be due to numerical
collisional effects.

\begin{figure}[!htb]
   \centering
   \includegraphics*[angle=0,width=210pt]{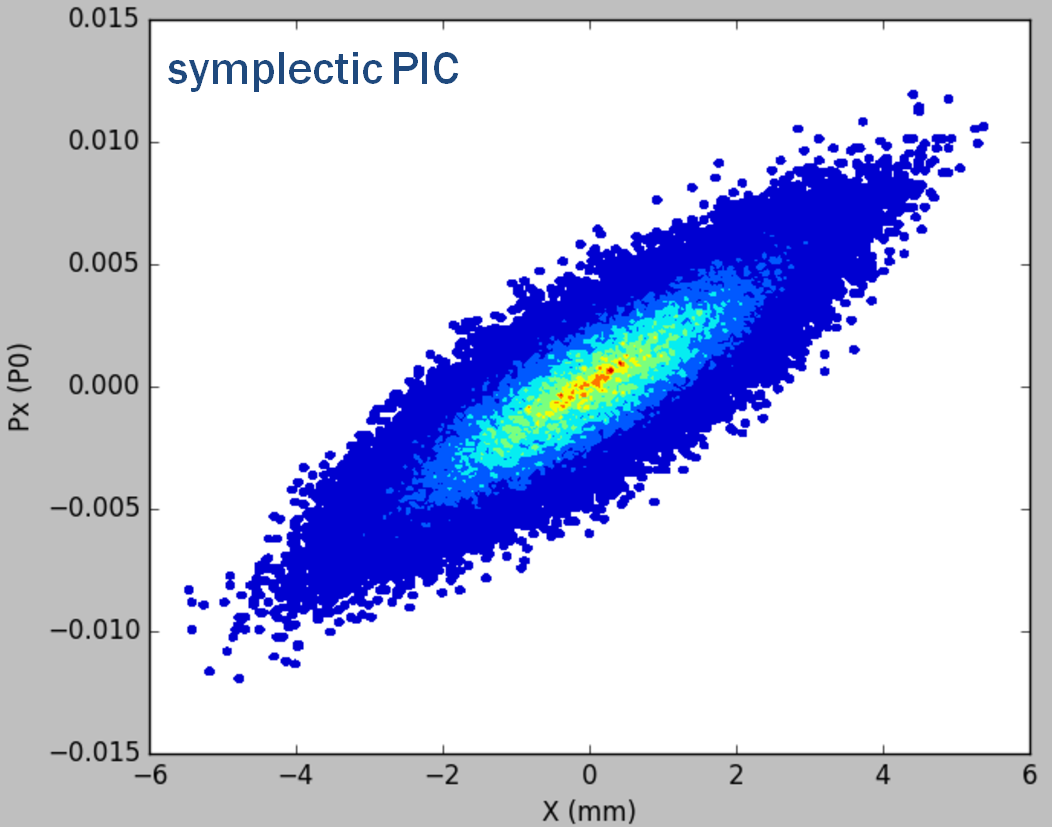}
   \includegraphics*[angle=0,width=210pt]{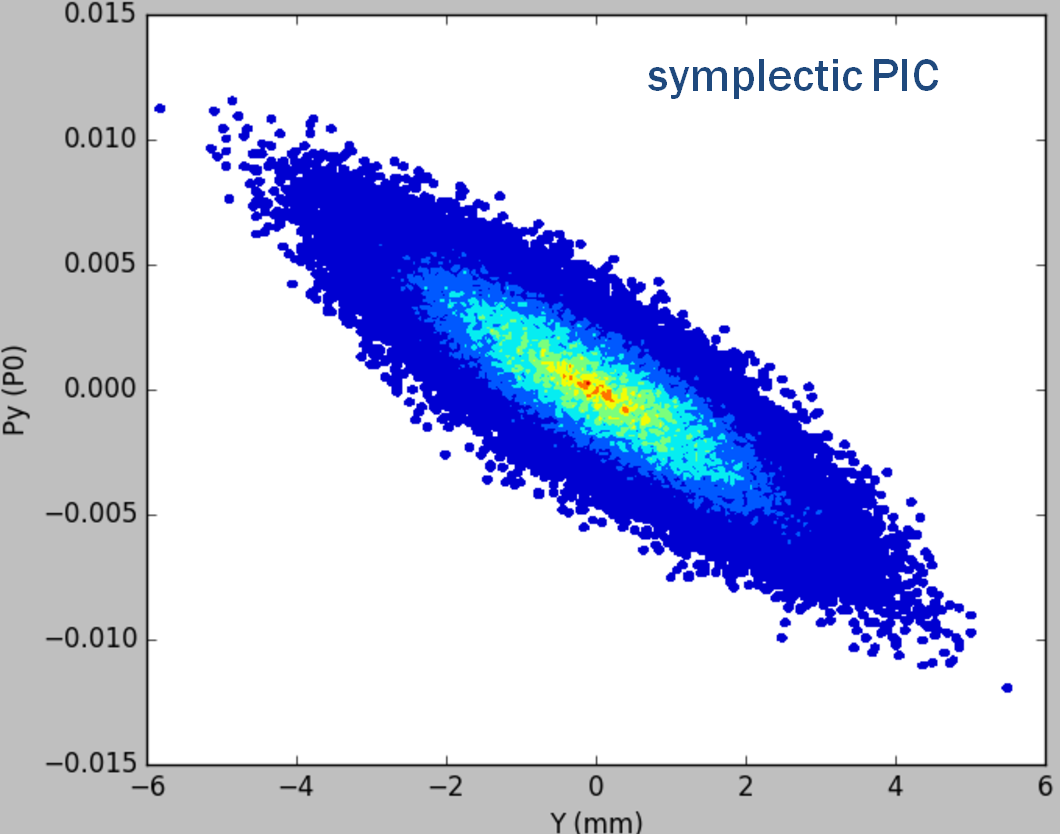}
   \includegraphics*[angle=0,width=210pt]{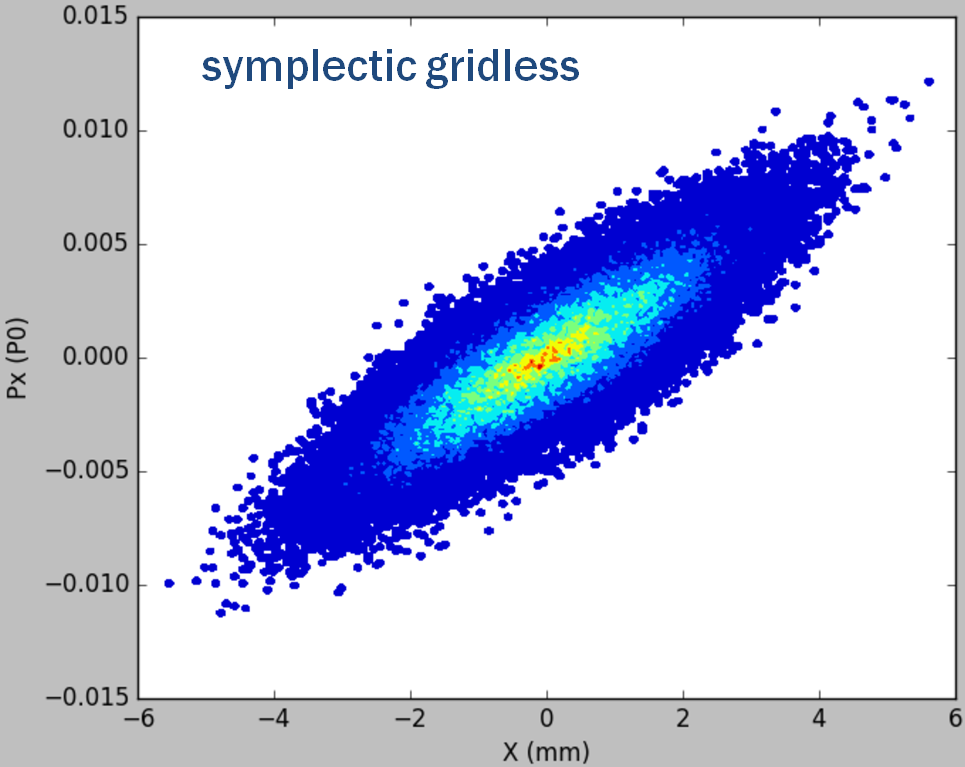}
   \includegraphics*[angle=0,width=210pt]{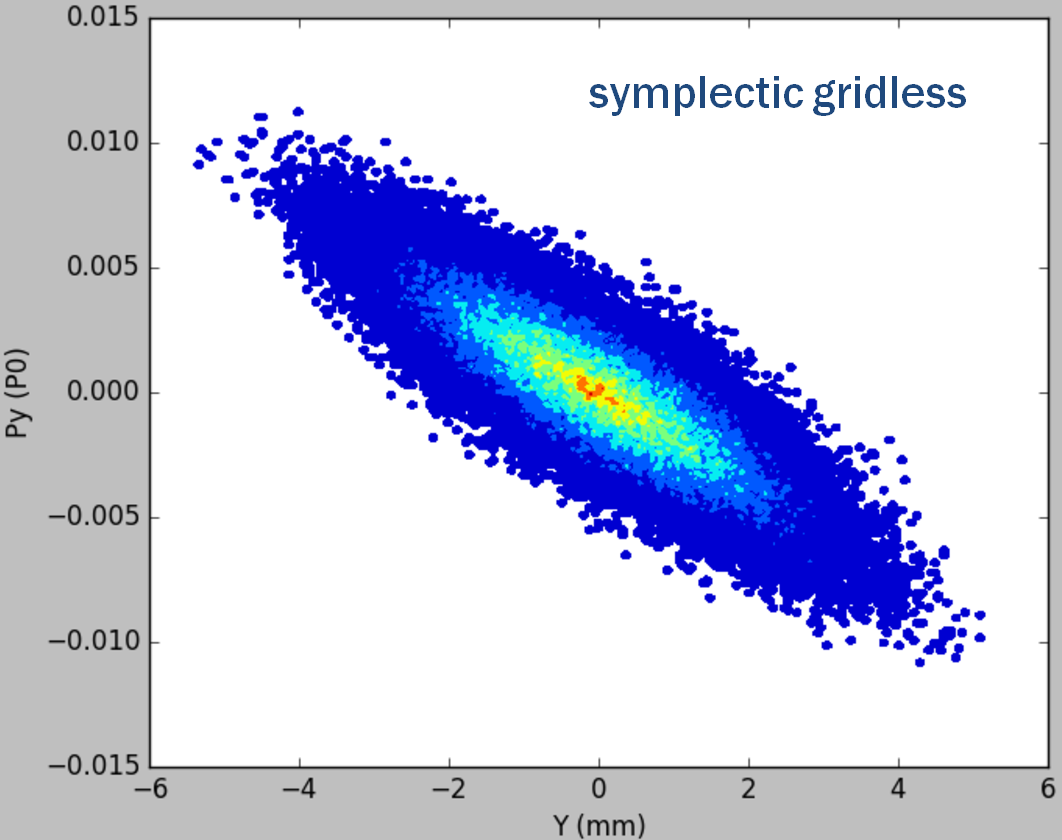}
   \includegraphics*[angle=0,width=210pt]{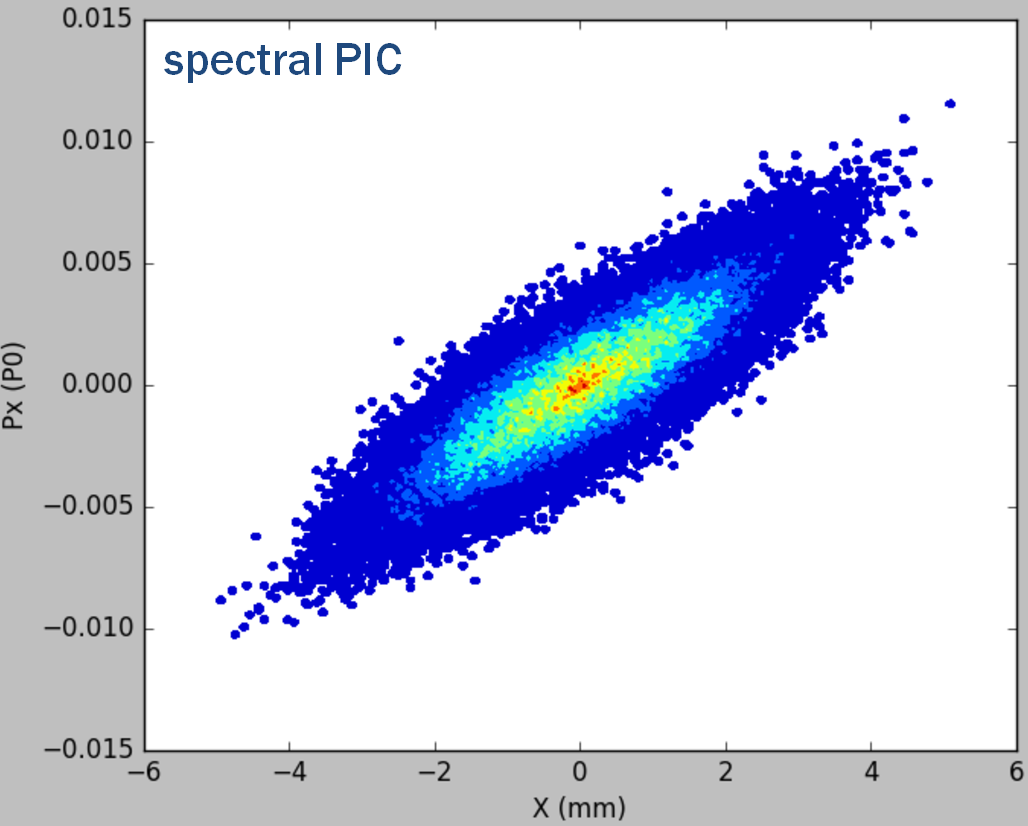}
   \includegraphics*[angle=0,width=210pt]{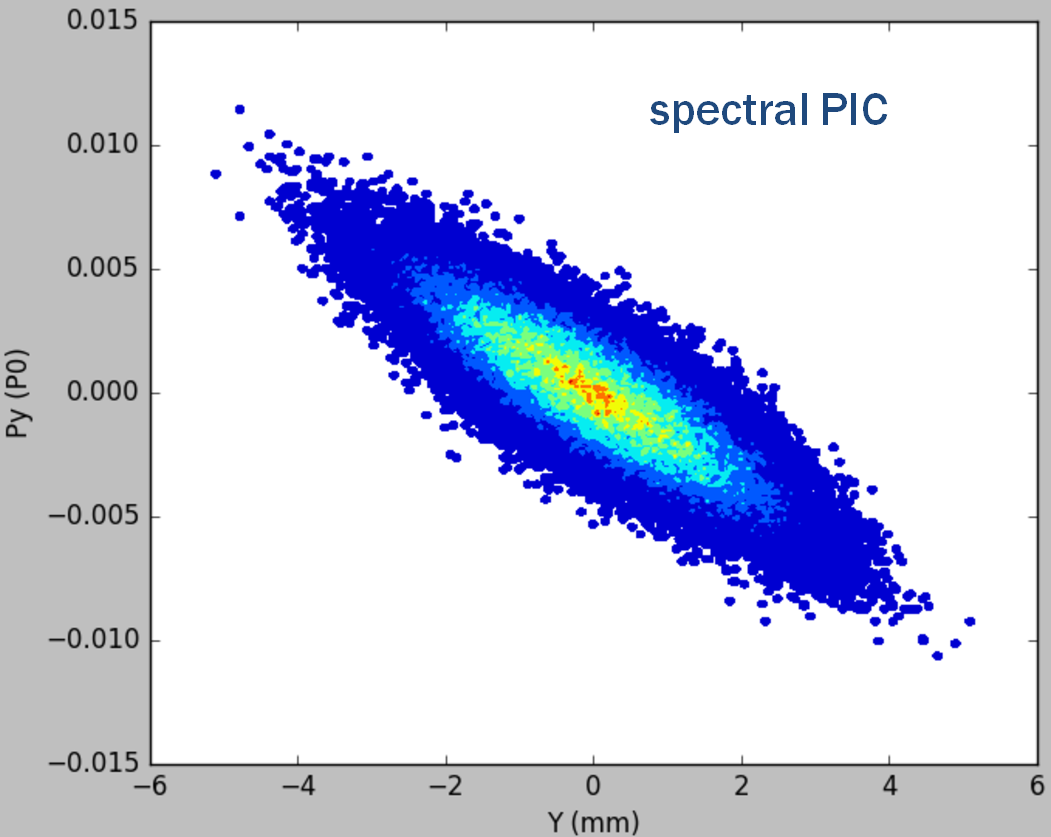}
	\caption{Final transverse phase space from the symplectic PIC 
	model (top), from the symplectic gridless particle model (middle),
	and from the nonsymplectic spectral PIC model (bottom).}
   \label{2dphasefodo}
\end{figure}
\begin{figure}[!htb]
   \centering
   \includegraphics*[angle=0,width=230pt]{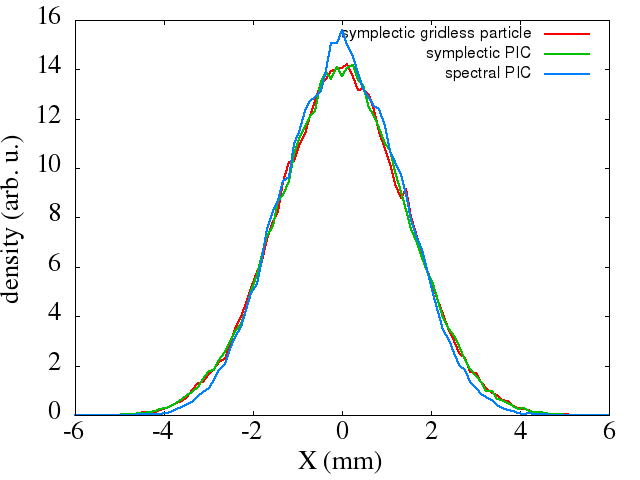}
   \includegraphics*[angle=0,width=230pt]{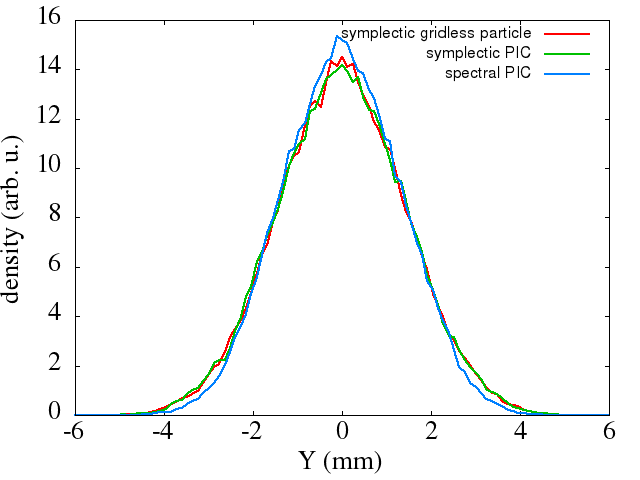}
   \caption{
	   Final horizontal (left) and vertical (right) density profiles
	  from the
	symplectic gridless particle model (red), the symplectic PIC model (green),
	and the nonsymplectic spectral PIC (blue).}
   \label{densfodo}
\end{figure}
\begin{figure}[!htb]
   \centering
   \includegraphics*[angle=0,width=230pt]{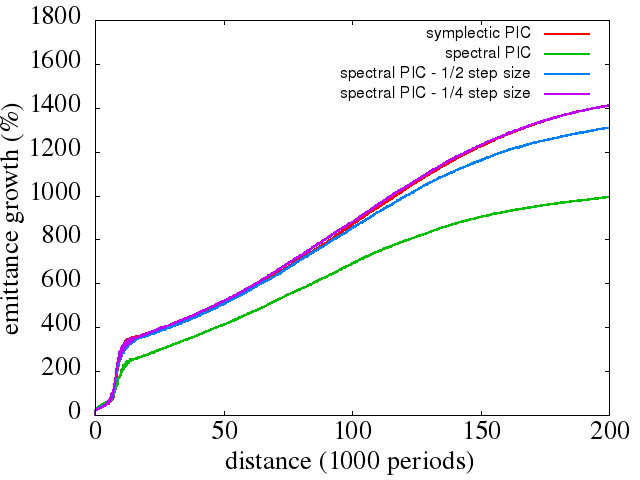}
   \caption{
	   Four dimensional emittance growth evolution from the
	symplectic symplectic PIC model (red), the nonsymplectic
	spectral PIC with the nominal step size (green), the spectral PIC
	with half of the nominal step size (blue), and the spectral PIC
	with one quarter of the nominal step size (pink).}
   \label{emtfodo2}
\end{figure}
Figure~\ref{2dphasefodo} shows the transverse phase space distributions
at the end of the simulation ($200,000$ periods) from these
three models. It is seen that the distributions from all three models
show similar phase space shapes after the strong collective resonance. 
However, the core density distributions show somewhat different structures.
The two symplectic models show similar structures while the nonsymplectic
spectral PIC model shows a denser core. 
This is also seen in the final horizontal
and vertical density profiles shown in Fig.~\ref{densfodo}.

The accuracy of the nonsymplectic PIC model can be improved with finer
step size. Figure~\ref{emtfodo2} shows the 4D emittance growth
evolution from the symplectic PIC model and those from the nonsymplectic PIC
model with the same nominal step size, from the nonsymplectic PIC model
with one-half of the nominal step size, and from the nonsymplectic PIC
model with one-quarter of the nominal step size. It is seen that as
the step size decreases, the emittance growth from the nonsymplectic PIC
model converges towards that from the symplectic PIC model.
\begin{figure}[!htb]
   \centering
   \includegraphics*[angle=0,width=230pt]{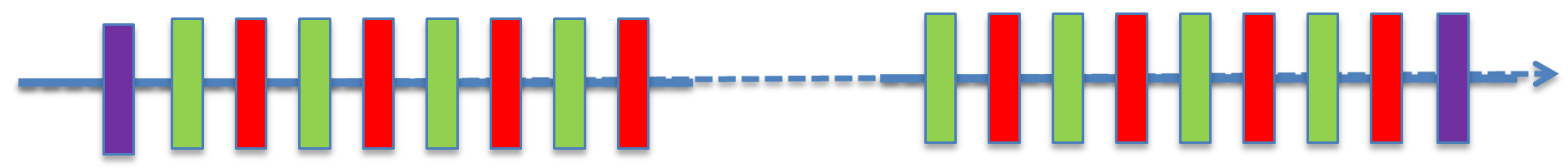}
   \caption{Schematic plot of a one-turn lattice that consists of $10$
	FODO lattice and one sextupole element.}
   \label{ringlat}
\end{figure}

In the second example, we simulated a $1$ GeV
coasting proton beam transport through a nonlinear lattice.
A schematic plot of the lattice is shown in Fig.~\ref{ringlat}.
Each period of nonlinear lattice consists of $10$ periods of $1$ m 
linear FODO lattice and a sextupole magnet to emulate a simple
storage ring.
The zero current horizontal and vertical tunes of the ring are $2.417$ and $2.417$
respectively. The integrated strength of the sextupole element is
$10$ T/m/m. The proton beam initial normalized transverse emittances are 
still $1$ um in both
directions with a 4D Gaussian distribution.
\begin{figure}[!htb]
   \centering
   \includegraphics*[angle=0,width=230pt]{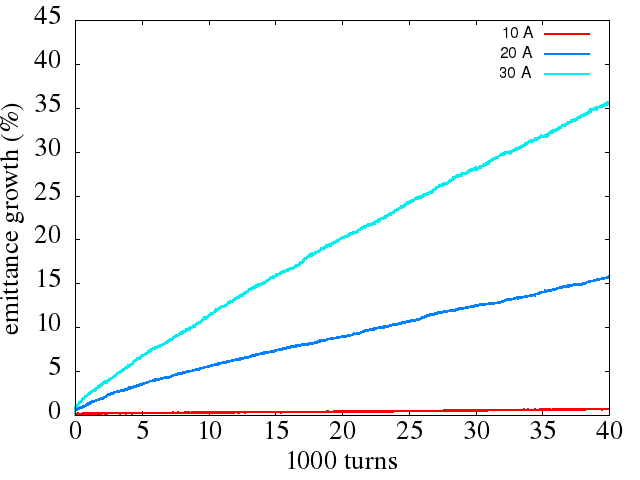}
   \includegraphics*[angle=0,width=230pt]{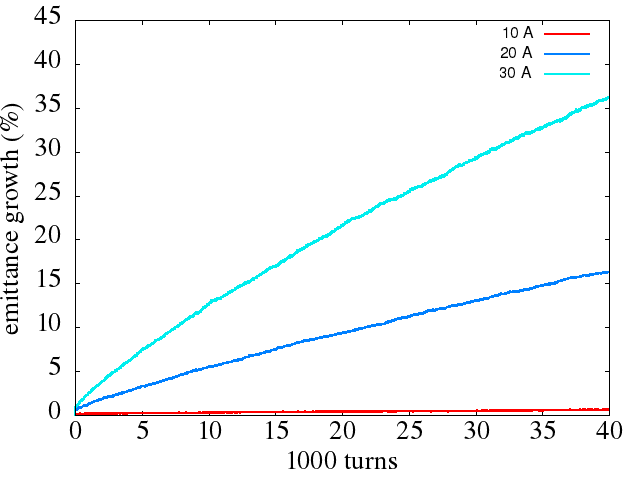}
   \includegraphics*[angle=0,width=230pt]{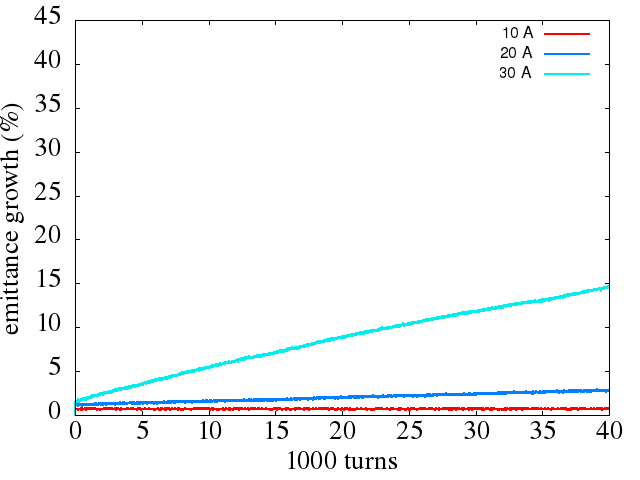}
   \caption{4D emittance growth evolution with $10$A, $20$A, and $30$A beam current
	from the symplectic PIC model (top), the symplectic gridless
	particle model (middle), and the nonsymplectic spectral PIC model 
	(bottom).}
   \label{emtring}
\end{figure}
\begin{figure}[!htb]
   \centering
   \includegraphics*[angle=0,width=210pt]{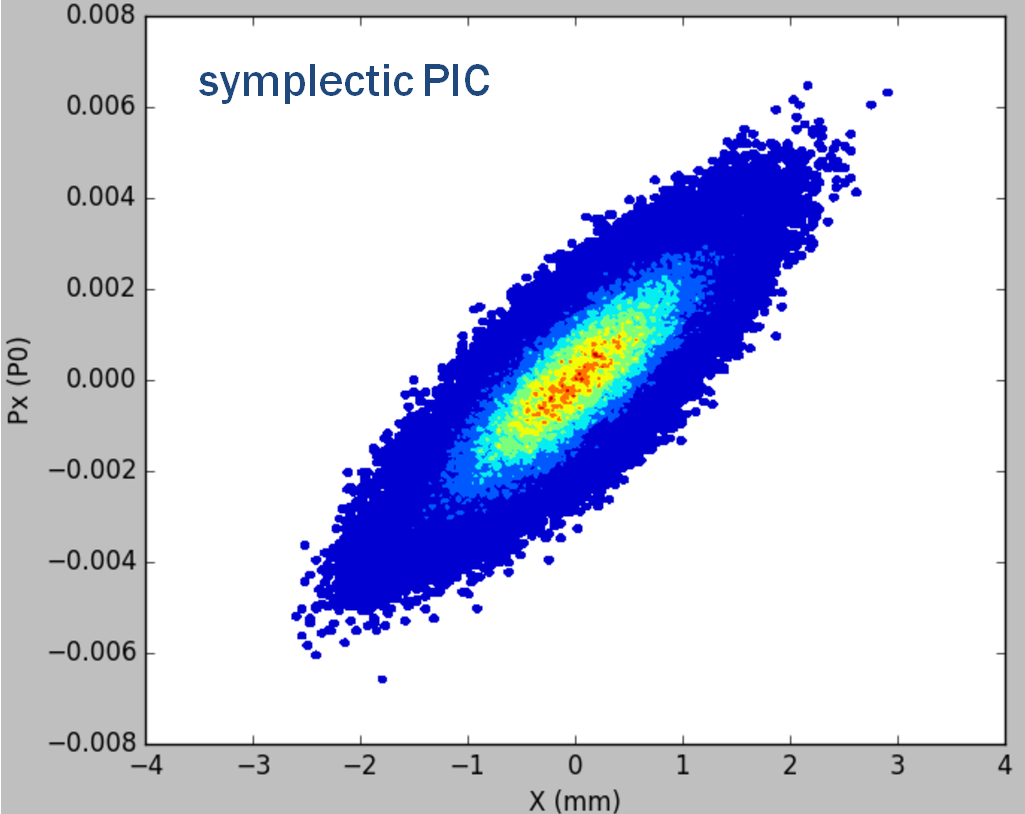}
   \includegraphics*[angle=0,width=210pt]{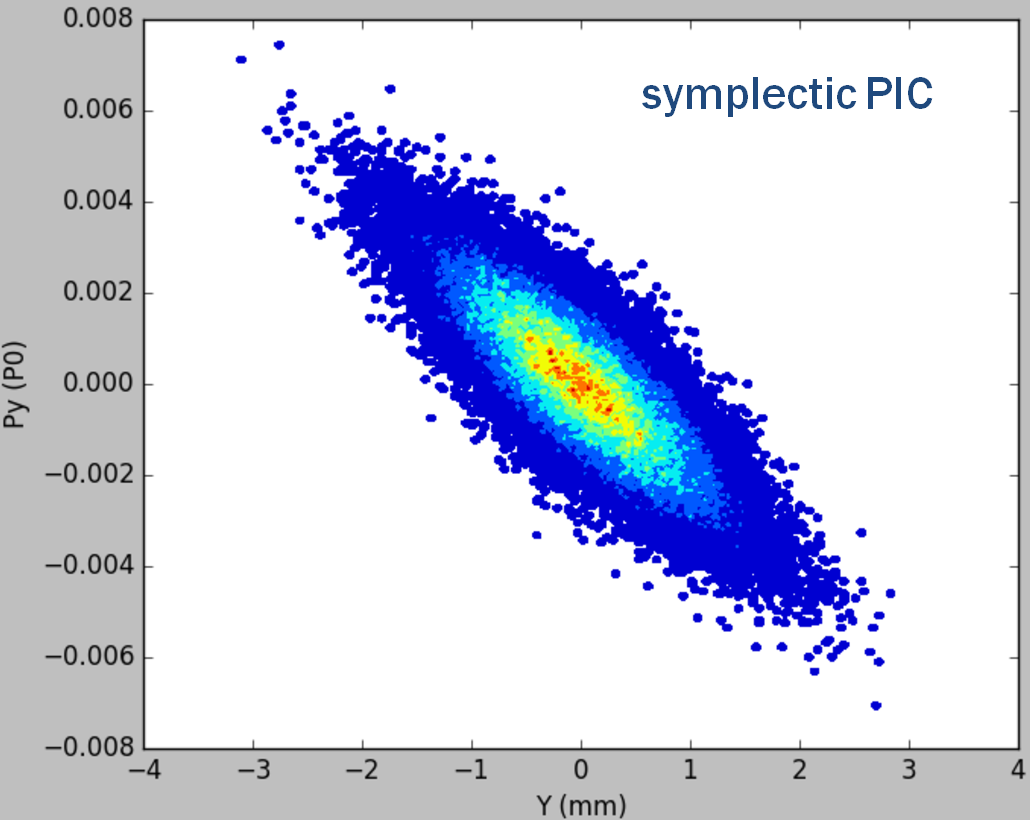}
   \includegraphics*[angle=0,width=210pt]{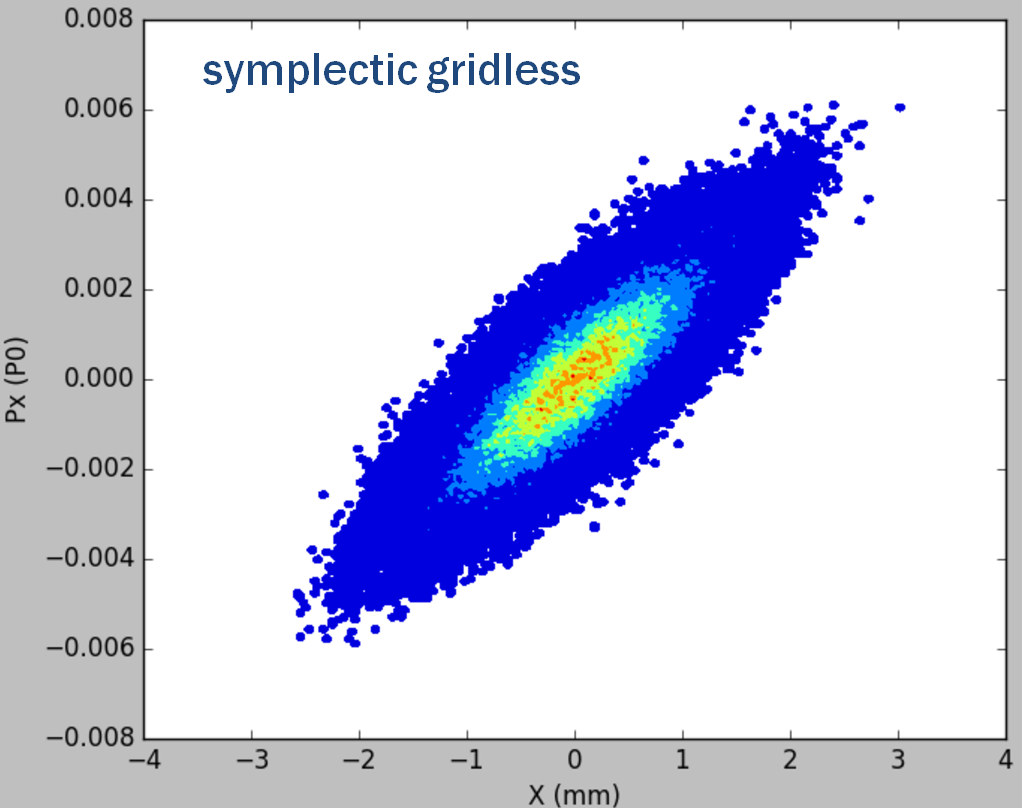}
   \includegraphics*[angle=0,width=210pt]{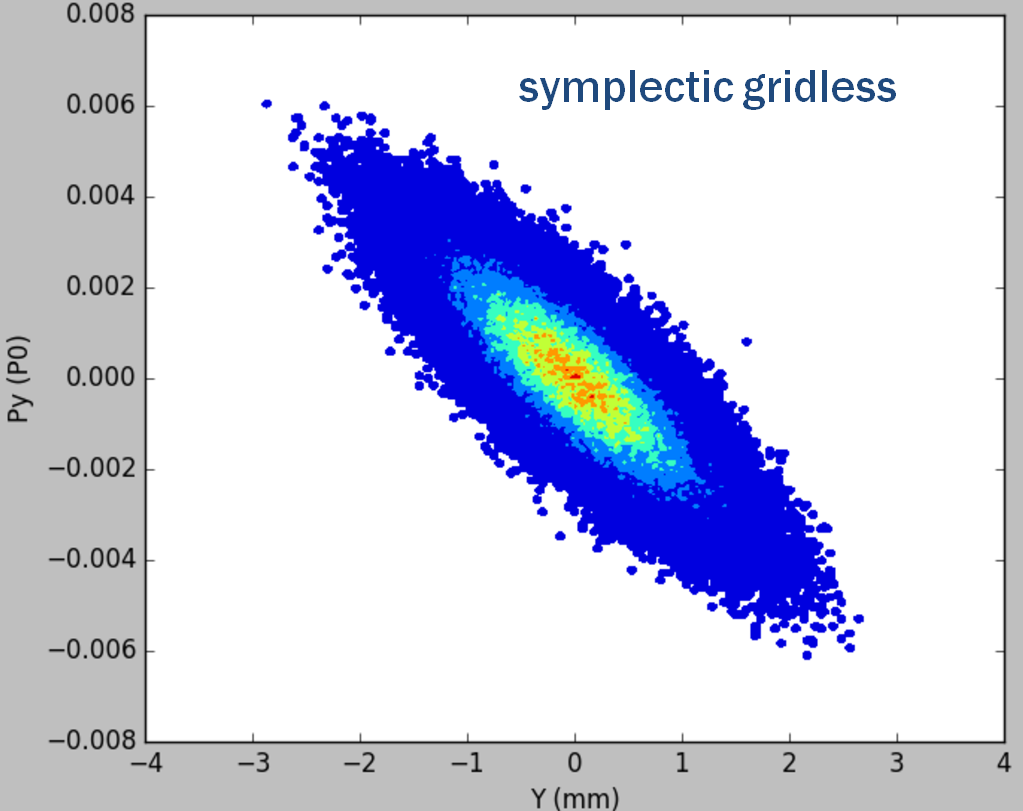}
   \includegraphics*[angle=0,width=210pt]{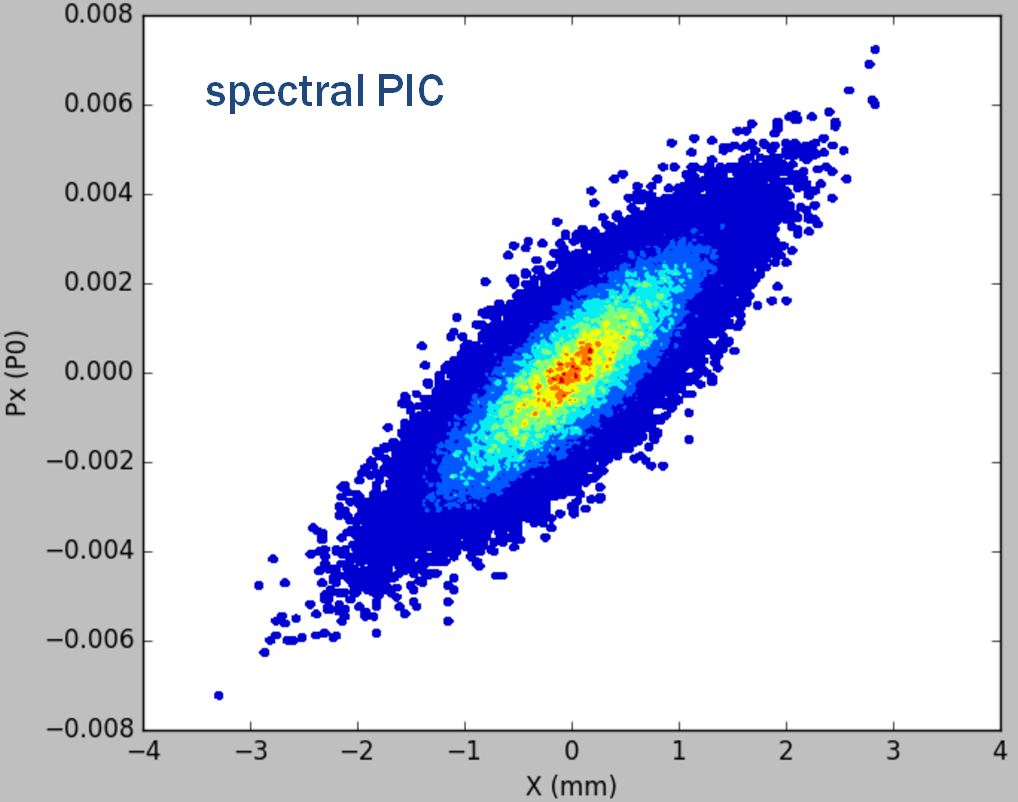}
   \includegraphics*[angle=0,width=210pt]{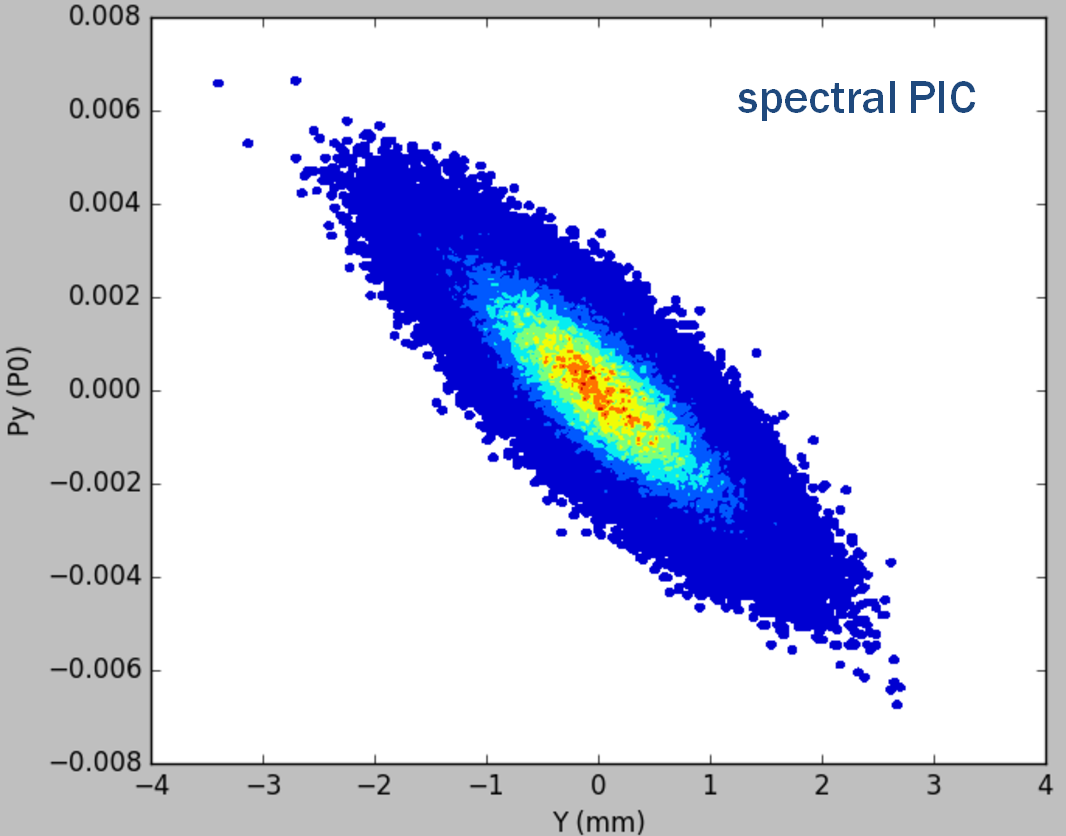}
   \caption{
	   Final transverse phase of the $30$A beam from the
	from the symplectic PIC model (top), the symplectic gridless
	particle model (middle), and the nonsymplectic spectral PIC model 
	(bottom).}
   \label{phasering}
\end{figure}
\begin{figure}[!htb]
   \centering
   \includegraphics*[angle=0,width=230pt]{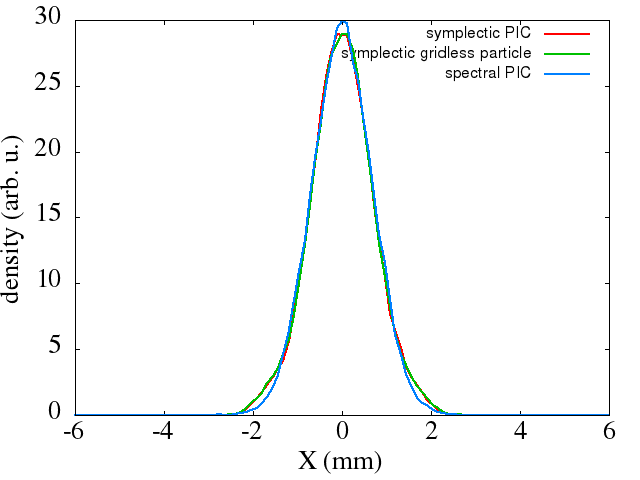}
   \includegraphics*[angle=0,width=230pt]{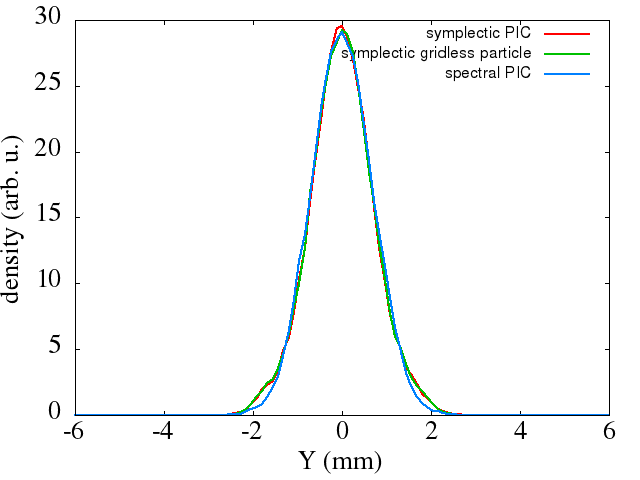}
   \caption{
	   Final horizontal (left) and vertical (right) density profiles
	  from the
	symplectic gridless particle model (red), the symplectic PIC model (green),
	and the nonsymplectic spectral PIC (blue).}
   \label{densring}
\end{figure}

Figure~\ref{emtring} shows the 4D emittance growth evolution with $10$ A, 
$20$ A, and $30$ A currents from the symplectic PIC model, the symplectic
gridless particle model, and the nonsymplectic PIC model.
The space-charge tune shift with $10$A beam current is $0.038$, with $20$ A is $0.075$,
and with $30$ A is $0.113$. 
With $10$ A current, particles stay out of the major $3^{rd}$
order resonance, 
there is little emittance growth. 
With $20$ A current, the proton beam moves near the $3^{rd}$ resonance,
some particles fall into the resonance and results in significant increase
of the emittance growth. With $30$ A current, a lot of particles fall into
the resonance and results in large emittance growth.
It is seen that from the above plot, all three models predict this
current dependent behavior 
qualitatively. The two symplectic models show very similar emittance
growth evolution for all three currents while the nonsymplectic 
PIC model shows much less emittance growth compared with the other two models.
This is probably due to the numerical damping effects 
in the nonsymplectic PIC model.

Figure~\ref{phasering} shows the final transverse phase space distributions
after $40,000$ turns. All three models show similar phase space shapes.
The two symplectic models show closer density distribution than the 
nonsymplectic PIC model. Figure~\ref{densring} shows the final 
horizontal and vertical density profiles from those three models.
The two symplectic models yield very close density profiles with
non-Gaussian halo tails.
The nonsymplectic PIC model yields denser core but less halo near the tail
density distribution.

\section{Conclusions}
In this paper, we proposed a symplectic PIC model and 
compared this model with the symplectic gridless particle model
and the conventional nonsymplectic PIC model for long-term
space-charge simulations of a coasting beam in a linear 
transport lattice and a nonlinear lattice.
Using the same step size and the same number of modes for space-charge solver,  
all three models qualitatively predict the similar emittance growth 
evolution due to the space-charge driven resonance
and yield similar final phase space shapes in the benchmark
examples.
Quantatively, the symplectic PIC and the symplectic
gridless particle model agree with each other very well, while
the nonsymplectic PIC model yields much less emittance growth
value.
Using finer step size, the emittance growth from the 
nonsymplectic PIC converges towards that from the symplectic PIC model. 

The computational complexity of the symplectic PIC model is proportional to
the $O(N_{grid}log(N_{grid})+N_p)$, where $N_{grid}$ and $N_p$ are total number
of computational grid points and macroparticles used in the simulation.
The computational complexity of the symplectic gridless particle model is
proportional to the $O(N_{mode}N_p)$, where $N_{mode}$ is the total number
of modes for the space-charge solver.
On a single processor computer, with a large number of macroparticles used,
the symplectic PIC model is computationally
more efficient than the symplectic gridless particle model.
However, on a massive parallel computer, the scalability of the
symplectic PIC model may be limited by the challenges of 
work load balance among multiple processors and communication associated
with the grided based space-charge solver and particle manager~\cite{qiangcpc}. 
The symplectic gridless particle model has regular data structure
for perfect work load balance and small
amount of communication associated with the space-charge solver.
It scales well on both multiple processor Central Processing Unit (CPU)
computers and multiple Graphic Processing Unit (GPU) computers~\cite{liu}.

\section*{ACKNOWLEDGEMENTS}
Work supported by the U.S. Department of Energy under Contract No. DE-AC02-05CH11231.
We would like to thank Drs. A. Friedman, C. Mitchell, R. D. Ryne, 
and J. Vay for discussions. This research used computer resources at the National Energy Research
Scientific Computing Center.

\end{document}